\documentclass[12pt]{iopart}

\usepackage[dvips]{graphicx}
\usepackage{iopams}
\usepackage{cite}

\expandafter\let\csname equation*\endcsname\relax
\expandafter\let\csname endequation*\endcsname\relax
\usepackage{amsmath}
\usepackage{amsmath}
\usepackage{mathrsfs}
\usepackage{xcolor}

 \DeclareMathOperator\arctanh{arctanh}

\begin{document}

\title{ Dissipation-induced enhancement of quantum fluctuations }
%
%

\author{Gianluca Rastelli}
\address{Fachbereich Physik, Universit{\"a}t Konstanz, D-78457 Konstanz, Germany}
\ead{gianluca.rastelli@uni-konstanz.de}

\begin{abstract}
We study a quantum harmonic oscillator linearly coupled through the position operator $\hat{q}$ to a first bath 
and through the momentum operator $\hat{p}$ to a second bath yielding an Ohmic-Drude dissipation.
We analyse the oscillator's fluctuations as a function of the ratio between the strength of the two couplings, 
focusing in particular on the situation in which the two dissipative interactions are comparable. 
Analytic formulas are derived in the relevant regimes corresponding to the low temperature limit and when 
the Drude high frequency cutoff  is much larger than all other frequencies. 
At low temperature, each bath operates to suppress the oscillator's ground state quantum fluctuations 
${\langle \Delta \hat{q}^2 \rangle}_0$ or ${\langle \Delta \hat{p}^2 \rangle}_0$ appearing in the corresponding interaction.
When one of the two dissipative interactions dominates over the other, the fluctuations for the coupling operator 
are  squeezed.
When the two interactions are comparable, the two baths enter in competition as the two conjugate operators do not commute 
yielding  {\sl quantum frustration}.  
In this regime, remarkably, the fluctuations of both two quadratures can be enhanced by increasing the dissipative coupling. 
\end{abstract}


%
%
%
%
%
%

\date{\today}

\maketitle

\section{Introduction}
\label{sec:intro}

The study of the quantum dissipation and the decoherence dynamics in atomic and mesoscopic systems 
is fueled by the perspective of engineering the reservoirs in order to preserve quantum coherence 
\cite{Weiss:2012,Breuer:2012,Wiseman:2014,Zagoskin:2011,Schlosshauer:2010}. 
This is the crucial point towards exploitable manipulation and control of individual quantum systems 
both for fundamental tests of quantum theory \cite{Leggett:2002,Raimond:2001,Haroche:2013,Wineland:2013,Bassi:2013,DeMartini:2012} 
and for the achievement of future quantum applications \cite{Nielsen:2012}.

The quantum harmonic oscillator is an exactly solvable reference system to understand quantum dissipation 
and decoherence \cite{Weiss:2012,Weiss:1985-2,Grabert:1988,Unruh:1989}. 
Moreover, many experimental coherent systems, for which quantum control is achievable or conceivable, are indeed harmonic oscillators.
These systems range from cavity Quantum ElectroDynamics \cite{Haroche:2013,Wineland:2013}  
to circuit microwave resonators \cite{Girvin:2009}, from electromechanical systems \cite{Poot:2012} to optomechanical systems \cite{Aspelmeyer:2014}
as well as other hybrid mesoscopic systems \cite{Xiang:2013,Dykman:2012}. 

For the quantum damped harmonic oscillator, it is known that the quantum fluctuations of the operator to which the bath is coupled are squeezed  
and those of its conjugate variable are enhanced in such a way that the Heisenberg uncertainty principle holds.
The case in which the oscillator is coupled to the bath through the position represents the standard conventional picture \cite{Weiss:2012}, 
whereas the case in which the oscillator is coupled to the bath through the momentum - which is the dual counterpart - is 
referred to in literature as unconventional or anomalous dissipation \cite{Cuccoli:2001,Ankerhold:2007}.

Remarkably, an open quantum system coupled to two independent environments by canonically conjugate operators shows an
enhancement of the quantum fluctuations. 
Moreover, the decay dynamics of decoherence and relaxation can be always underdamped despite the fact that the 
strength of the dissipative interaction increases. 
This state of affairs was termed ``quantum frustration'' and was analysed for an open quantum system realized by a harmonic 
oscillator \cite{Kohler:2005,Kohler:2006,Cuccoli:2010,Kohler:2013-2} or a single spin \cite{Neto:2003,Novais:2005,Kohler:2013,Bruognolo:2014,Zhou:2015}.
These findings can be understood by considering the two baths as two detectors continuously coupled to the system and measuring  
simultaneously two non-commutating observables.
This frustration of decoherence and dissipation can be attributed to the noncommuting nature of the conjugate coupling 
operators that prevents the selection of an appropriate pointer basis to which the quantum system could relax.
Quantum frustration due to competing dissipative processes has also been studied for a many-spin system \cite{Lang:2015}.

In this work, we consider a symmetric environmental coupling for the position $\hat{q}$ and for the momentum $\hat{p}$ of a quantum 
harmonic oscillator.
We focus on the case of ohmic dissipation with a Drude large frequency cut-off. 
The phase diagram presents regions where the system shows an enhancement or a squeezing of the quantum fluctuations.
Compared to the previous works \cite{Kohler:2006,Cuccoli:2010}, here we derive analytical formulas which allow to analyze  
in detail these effects and the role of the temperature as well as of the high frequency cut-off of the baths' spectrum.
The analytic results show that such quantum fluctuations (squeezed or enhanced) are observable at low temperatures $T < T^*$, 
where $T^*$ is the typical temperature below which finite temperature corrections are negligible and the fluctuations of the particle 
are controlled by the quantum contribution.
Analytic results also point out that quantum fluctuations exhibit  an universal contribution - independent of the large-frequency cutoff $\omega_c$ - 
and a part which scales logarithmically with $\omega_c$.

The paper is organized as follows:  in section \ref{sec:model_2bath}, we introduce the model Hamiltonian for the quantum harmonic oscillator coupled to two baths and derive the expressions for the fluctuations of $q$ and $p$.
In section \ref{sec:Ohmic_diss}, we characterize the environment interaction and provide an analytic expression for the fluctuations 
which allows the analytic expansion of the fluctuations in high and low temperature regime discussed in section \ref{sec:Hi_t}.
Hence, in section \ref{sec:T=0}, we focus on the analysis of the zero temperature fluctuations.
A short summary and perspective are given in the last section \ref{sec:conclusions}.

\section{Dissipative interaction with two independent baths}
\label{sec:model_2bath}

The Hamiltonian of the harmonic oscillator linearly coupled to two baths is
%
%
%
%
%
\begin{equation}
\hat{H} = \frac{\mbox{}\hat{p}^2}{2m} +   \frac{m\omega_0^2}{2} \hat{q}^2 + \hat{H}_{q}+ \hat{H}_{p} \, , 
\end{equation}
%
%
%
%
%
with the two conjugates operators $[\hat{q},\hat{p}] = i\hbar$ and the interaction with the environment 
described by two ensembles of independent harmonic oscillators 
%
%
%
%
%
\begin{subequations}
\begin{align}
\hat{H}_{q}   &= \sum_n 
\left[ 
\frac{ \hat{P}^2_{q,n}}{2m_{q,n}} 
+   
\frac{1}{2} m_{q,n}^{\phantom{g}}\omega_{q,n}^2
{\left( \hat{Q}_{q,n}   -  \frac{\lambda_{q,n} \hat{q}}{ m_{q,n}^{\phantom{2}} \omega_{q,n}^2}   \right)}^2 
\right] ,   \label{q:H_env_q}  \\
\hat{H}_{p}  &= \sum_n  \left[ 
\frac{1}{2m_{p,n}}  {\left( \hat{P}_{p,n}   - \frac{\lambda_{p,n}\hat{p}}{m\omega_{p,n}\omega_0} \right)}^2 
+
\frac{1}{2} m_{p,n}^{\phantom{g}}\omega_{p,n}^2   \hat{Q}^2_{p,n}
\right]  \, , \label{eq:H_env_p}
\end{align}
\end{subequations}
%
%
%
%
%
%
%
with the  conjugates operators $[\hat{Q}_{\nu,n},\hat{P}_{\nu,n'}] = \, i\hbar \delta_{\nu,\nu'} \delta_{n,n'}$
where $\nu=q,p$ are the two bath indices.
For the interacting Hamiltonians $\hat{H}_{q}$ and $\hat{H}_{p}$ the coupling constants ${\lambda_{\nu,n}}$ have the same dimensions.
Using the equations of motion in the Heisenberg picture $\hat{O}(t)=e^{iHt /\hbar}\hat{O}e^{-iHt /\hbar}$, we obtain 
%
%
%
%
%
\begin{subequations}
\begin{align}
\frac{d\hat{p}(t)}{dt} & =  -m \omega_0^2 \hat{q}(t)+ \hat{F}_q(t) - \int^{+\infty}_{t_0} \!\!\!\!\!\! \!\! dt' \, \eta_q(t-t') \, m \frac{d \hat{q}(t')}{dt'}  \, , \label{eq:Langevin_1} \\
m \frac{d^2\hat{q}(t)}{dt^2} & = \frac{d\hat{p}(t)}{dt}+ \hat{F}_p(t) + \int^{+\infty}_{t_0} \!\!\!\!\!\! \!\! dt' \, \eta_p(t-t') \, \frac{1}{\omega_0^2}\frac{d^2 \hat{p}(t')}{dt'^2} \, , 
\label{eq:Langevin_2}
\end{align}
\label{eq:Langevin}
\end{subequations}
in which we have introduced $t_0$ as the initial time for the interaction and the two response functions of the two baths as
%
%
%
%
\begin{equation}
\label{eq:eta_t}
\eta_{\nu}(t) = \frac{\theta(t)}{m} \sum_{n} \frac{\lambda_{\nu,n}^2}{m_{\nu,n}^{\phantom{2}}\omega_{\nu,n}^2} \cos(\omega_{\nu,n}t) \, .
\end{equation}
The two response functions $\eta_{\nu}(t)$  satisfy causality and the Kramers-Kronig relations.
For times $t>t_0$, the two force operators describing the quantum noise read 
%
%
%
%
%
\begin{subequations}
\begin{align}
\hat{F}_q(t) &=  \sum_{n} \lambda_{q,n} \hat{Q}_{q,n}^{(0)}(t-t_0)  - \eta_q(t-t_0) m \hat{q}_{t_0} \, , \\
\hat{F}_p(t) &=  \sum_{n} \lambda_{p,n} \frac{ \omega_{p,n} }{\omega_0} \hat{Q}_{p,n} ^{(0)}(t-t_0)  + \eta_p(t-t_0)
\frac{1}{m\omega_0^2} {\left. \frac{d\hat{p}(t)}{dt} \right|}_{t=t_0} \, ,
\end{align}
\end{subequations}
in which ${\hat{Q}_{\nu,n}^{(0)}(t-t_0)}$  are the free evolution operators for the two baths 
and $\hat{q}_{t_0}$ and ${\left. d\hat{p}/dt \right|}_{t_0} $ are the oscillator's position operator and time derivative of the momentum 
at the initial time $t_0$. 
The functions $\eta_{\nu}(t) $ are characterized by a typical correlation time $\sim 2\pi/\omega_c$ 
with $\omega_c$ being a large-frequency cutoff.
Thus  we assume that the response functions vanish as $ \eta_q(t-t_0) \simeq 0$ for long times $\omega_c(t-t_0)\gg1$.
Hence, the operators $\hat{F}_{q}$ and $\hat{F}_{p}$ reduce to
%
%
%
%
%
%
\begin{subequations}
\begin{align}
\hat{F}_q(t) &\simeq
\sum_{n} \sqrt{ \frac{\hbar  \lambda_{q,n}^2}{2m_{q,n}\omega_{q,n}} }  \left( \hat{a}_{q,n}e^{-i\omega_{q,n}(t-t_0)}+  \mbox{h.c.}  \right) 
\label{eq:Forces_longtime1} \\
\hat{F}_{p}(t) &\simeq
\sum_{n} \sqrt{ \frac{\hbar  \lambda_{p,n}^2}{2m_{p,n}\omega_{p,n}} }  \left( \frac{\omega_{p,n}}{\omega_0} \right)
\left( \hat{a}_{p,n}e^{-i\omega_{\nu,n}(t-t_0)}+  \mbox{h.c.}   \right)    \label{eq:Forces_longtime2} 
\end{align}
\label{eq:Forces_longtime}
\end{subequations}
in which we used the creation and annihilation operator for both baths as 
$\hat{Q}_{\nu,n}^{(0)}=\sqrt{\hbar/(2m_{\nu,n}\omega_{\nu,n})} (\hat{a}_{\nu,n} + \hat{a}_{\nu,n}^{\dagger})$. 
For the initial state, we assume the total density matrix factorized as $\rho_{t_0}= \rho_0 \rho_q\rho_p $
with $\rho_0$ the initial state of the oscillator, $\rho_{q,p}$ the thermal density matrices for the two baths 
$\rho_{\nu} \propto \exp(-\beta \sum_n \omega_{\nu,n} \hat{a}_{\nu,n}^{\dagger} \hat{a}_{\nu,n}^{\phantom{g}})$ 
and $\beta=\hbar/(k_BT)$. 
Then the correlation functions of the noise operators are time translational invariant.
From Eqs.~(\ref{eq:Forces_longtime}) and using the Fourier transform $\hat{F}(\omega)=\int\!\!dt \exp(-i\omega t) \hat{F}(t)$ we 
obtain 
%
%
%
%
%
%
%
\begin{subequations}
\begin{align}
\langle \hat{F}_{q}(\omega_1) \hat{F}_{q}(\omega_2)  \rangle  &= {(2\pi)}^2 \delta(\omega_1+\omega_2) S_{q}(\omega_1)  \, , \label{eq:F_t1t2} \\
\langle \hat{F}_{p}(\omega_1) \hat{F}_{p}(\omega_2)  \rangle & = {(2\pi)}^2 \delta(\omega_1+\omega_2)  {(\omega/\omega_0)}^2   S_{p}(\omega_1)  \, , 
\end{align}
\end{subequations}
where we introduced the noise spectral function
%
%
%
%
%
%
%
\begin{equation}
\label{eq:S_omega}
S_{\nu}(\omega)  \! = \!
\sum_{n} \hbar \omega_{\nu,n}  \left( \frac{\lambda_{\nu,n}^2}{2 m_{\nu,n}^{\phantom{2}} \omega_{\nu,n}^2}  \right)
\left[ (n_B(\omega_{\nu,n})+1) \delta(\omega+\omega_{\nu,n}) + n_B(\omega_{\nu,n}) \delta(\omega-\omega_{\nu,n}) 
\right] \, ,
\end{equation}
with the Bose factor $n_B(\omega)=1/(e^{\beta\omega}-1)$.
The noise spectral function can be related to the response function of the baths 
via 
%
%
%
%
%
%
\begin{equation}
\mbox{Re}[\eta_{\nu}(\omega)]= 
\frac{1}{m} \sum_n \frac{\pi \lambda_{\nu,n}^2}{2m_{\nu,n}^{\phantom{2}}\omega_{\nu,n}^2} 
\left[ \delta(\omega-\omega_{\nu,n})  +   \delta(\omega+\omega_{\nu,n})  \right]  \, ,
\end{equation}
which follows from the definition of $\eta(t)$ in Eq.~(\ref{eq:eta_t}). 
Assuming $t_0\rightarrow-\infty$, the Eqs.~(\ref{eq:Langevin}) can be solved using the Fourier transform.
The results read
%
%
%
%
%
%
\begin{equation}
\label{eq:solution}
\left[
\begin{array}{c}
\hat{q}(t)   \\
\hat{p}(t)  
\end{array}
\right]
=  \frac{1}{2\pi} \int^{+\infty}_{-\infty}\!\!\!\!\!\!\!\! d\omega  \,
e^{i \omega t}
\frac{1}{D(\omega)}
\left[
\!\!\!\!
\begin{array}{cc}
\frac{1}{m}(1+i\frac{\omega}{\omega_0^2}\eta_p(\omega))  & \frac{1}{m}  \\
i\omega 										& i \frac{\omega_0^2}{\omega}(1+i\frac{\omega}{\omega_0^2}\eta_q(\omega))\\
\end{array}
\!\!\!\!
\right]
\left[
\begin{array}{c}
\hat{F}_q(\omega)   \\
\hat{F}_p(\omega)  
\end{array}
\right] \, , 
\end{equation}
with
%
%
%
%
%
%
\begin{equation}
\label{eq:Den}
D(\omega)=\omega_0^2-\omega^2+i\omega[\eta_q(\omega)+\eta_p(\omega)]-\frac{\omega^2}{\omega_0^2}\eta_q(\omega)\eta_p(\omega) \, . 
\end{equation}
From Eq.~(\ref{eq:solution}) it is possible to compute the correlation functions of the oscillators for arbitrary products of the position and momentum.
We now focus our attention on the two fluctuations.
After some algebra, we obtain
%
%
%
%
%
%
\begin{equation}
\label{eq:q2}
\left[
\begin{array}{c}
\langle\hat{q}^2\rangle/q_0^2  \\
\langle\hat{p}^2\rangle/p_0^2 
\end{array}
\right]
= -\frac{1}{\pi \omega_0}
\int^{+\infty}_{-\infty}\!\!\!\!\!\!\!\!\! d\omega \, \coth(\beta\omega/2) \, 
 \mbox{Im}\left(
\frac{1} {D(\omega)}
\left[
\begin{array}{c}
\omega_0^2+i\omega\eta_p(\omega) \\
\omega_0^2+i\omega\eta_q(\omega)
\end{array}
\right]
\right)
\end{equation}
with the normalizations $q_0^2=\hbar/(2m\omega_0)$ and $p_0^2=m\hbar\omega_0/2$.
Provided that the poles of the functions $D(z)$ - with $z$ complex - have always the same 
sign for the imaginary part, then we can calculate the integral using the residues theorem for a 
closed curve lying only in one half of the complex plane which  contains only the poles of function $\coth(\beta\omega/2)$.
The latter correspond to the Matsubara frequencies $z_k = i \omega_k =i 2\pi k/\beta $ with $k$ integer.
This yields 
%
%
%
%
%
%
\begin{equation}
\label{eq:q2_sum}
\left[
\begin{array}{c}
\langle\hat{Q}^2\rangle  \\
\langle\hat{P}^2\rangle 
\end{array}
\right]
= 
\left[
\begin{array}{c}
\langle\hat{q}^2\rangle/q_0^2  \\
\langle\hat{p}^2\rangle/p_0^2 
\end{array}
\right]
= \frac{2}{\beta\omega_0}+ 
\frac{4}{\beta\omega_0}
\sum_{k=1}^{+\infty}
\frac{1} {D(-i\omega_k)}
\left[
\begin{array}{c}
\omega_0^2+ \omega_k \eta_p(-i\omega_k) \\
\omega_0^2+\omega_k \eta_q(-i\omega_k)
\end{array}
\right]  \, 
\end{equation}
and the Heisenberg  uncertainty relation for the scaled operators $\hat{Q}/q_0$ and $\hat{P}/P_0$ read  
$\langle \hat{Q}^2 \rangle \langle \hat{P}^2 \rangle  \geq 1$.
Eq.~(\ref{eq:q2_sum}) is for a harmonic oscillator linearly coupled to two independent 
baths with arbitrary dissipative interactions.
The  formulas for the two quadratures are symmetric under interchange of the two response functions    
$\eta_{q} \leftrightarrow \eta_{p}$, i.e.
%
%
%
%
%
%
%
%
\begin{equation}
\label{eq:quadratures}
\langle \hat{Q}^2 \rangle =  \sigma(\eta_q,\eta_p) \equiv \sigma_{q} \, , \quad
\langle \hat{P}^2 \rangle =  \sigma(\eta_p,\eta_q) \equiv  \sigma_{p} \, .
\end{equation}
Due to this symmetry, hereafter we discuss the function $\sigma_{q}$. 
Finally, I point out that the result (\ref{eq:q2_sum}) can be also obtained by using  the path integral.
This confirms the initial assumption that the poles of the functions $D(z)$ have always the same 
sign for the imaginary part.
An explicit demonstration is discussed in the next sections.

\section{The Ohmic-Drude dissipation}
\label{sec:Ohmic_diss}

Here we focus on the case in which the oscillator is coupled to the two baths via 
an ohmic dissipation with a Drude large frequency cutoff $\omega_c$.
For this case the two response functions read
%
%
%
%
%
\begin{equation}
\eta_{\nu}(t) =  \theta(t) \,\, \gamma_{\nu} \,\, \omega_c e^{-\omega_c t}
\, , \quad
\eta_{\nu}(\omega) = \gamma_{\nu}/(1+i \omega/\omega_c)
\, ,
\end{equation}
in which $\gamma_{\nu}$ are the damping coefficients (with dimensions of a frequency).
Notice that, indeed,  the function $\eta_{\nu}(t) \rightarrow 0$ for large times $\omega_c t \gg 1$ as assumed  
in the previous section.
Then the formulas (\ref{eq:q2_sum}) and (\ref{eq:quadratures}) can be simplified to
%
%
%
%
%
%
\begin{equation}
\label{eq:q2_sum_ohmic}
\sigma_{q} = \frac{2}{\beta\omega_0} +
\frac{4}{\beta\omega_0}  \sum_{k=1}^{+\infty}
\frac{ \omega_0^2{(\omega_c+\omega_k)}^2 + \gamma_p \omega_k \omega_c  (\omega_c+\omega_k)
}{ (\omega_k^2 +\omega_0^2){(\omega_c+\omega_k)}^2 + (\gamma_q+\gamma_p) \omega_k\omega_c{(\omega_c+\omega_k)}
+\omega_k^2 ( \frac{\gamma_q\gamma_p}{\omega_0^2} ) \omega_c^2} \, .
\end{equation}
This result is in agreement with Ref.\cite{Cuccoli:2010}, where the function $\sigma_q$ was determined numerically 
and the results were discussed at vanishing temperature. 
In this work we proceed in a way similarly to the case of a damped harmonic oscillator with a single bath \cite{Weiss:2012}. 
We note that the sum over the Matsubara frequencies  Eq.~(\ref{eq:q2_sum_ohmic}) 
can be carried out analytically if one introduces the frequencies ${\Omega_i}$ as 
the negative roots of the quartic polynomial in $\omega_n$ in the denominator.
They are defined as 
%
%
%
%
%
%
%
\begin{equation}
\label{eq:P4}  
\prod_{i=1}^{4}  (\omega_k+\Omega_i)  =
(\omega_k^2 +\omega_0^2){(\omega_c+\omega_k)}^2 
+ (\gamma_q+\gamma_p) \omega_k\omega_c{(\omega_c+\omega_k)}
+ \omega_k^2 \left( \frac{\gamma_q\gamma_p}{\omega_0^2} \right) \omega_c^2  
\end{equation}
and satisfy the relations: $\sum_i \Omega_i = 2\omega_c$,  
$\sum_{i \neq j} \Omega_i \Omega_j = 2[\omega_c^2 + \omega_0^2 + \omega_c(\gamma_q+\gamma_p)]$, 
$\sum_{i \neq j \neq k} \Omega_i \Omega_j\Omega_k = 3 \omega_c [2 \omega_0^2 + \omega_c(\gamma_q+\gamma_p)]$ 
and $ \Omega_1 \Omega_2  \Omega_3  \Omega_4 = \omega_c^2 \omega_0^2$.
In this way, we obtain 
%
%
%
%
%
%
%
\begin{equation}
\label{eq:q2_final}  
\sigma_{q} = \frac{2}{\beta\omega_0} +
\frac{2\omega_0}{\pi\omega_c} \,
\sum_{i=1}^{4} \, A_i \, \Psi \left(1 +  \frac{\beta\Omega_i}{2\pi}  \right)
\, ,
\end{equation}
in which $\Psi$ is the digamma function and the coefficients ${A_i}$ are given by
%
%
%
%
%
%
\begin{equation}
\label{eq:A_coeff}
A_i = 
\frac{\omega_c (\omega_c-\Omega_i)
\left[\omega_c-\Omega_i(1+\gamma_p\omega_c/\omega_0^2)\right]}{(\Omega_i-\Omega_j)(\Omega_i-\Omega_k)(\Omega_i-\Omega_{\ell})}  \, ,
\quad 
\mbox{for}
\quad (j,k,\ell) \neq i  \, ,
\end{equation}
with $i,j,k,\ell=1,2,3,4$.
To conclude this section, we note that Eq.~(\ref{eq:q2_final}) represents one the main results 
of this work, encoding the quantum fluctuations of a harmonic oscillator coupled to two different 
baths via the two conjugate variables $\hat{q},\hat{p}$ at arbitrary temperature and frequency 
cutoff $\omega_c$  for the Ohmic-Drude dissipation.
This analytic expression allows to investigate the physical behavior in the different parameter regimes. 
In particular, we will now discuss the enhancement of the fluctuations,  the role of the temperature $T$ as 
well as of the large frequency cutoff $\omega_c$ in the spectrum of the baths.

\section{High and low temperature limits}
\label{sec:Hi_t}

First we discuss the behavior of Eq.(\ref{eq:q2_final}) at finite temperature.
At high temperature, we recover the classical limit. 
More precisely, for sufficiently high temperature, such that $\beta \Omega_i /(2\pi) \ll 1$ for $i=1,\dots,4$, we find the result of the  
equipartition theorem:
%
%
%
%
%
%
\begin{equation}
\label{eq:classica}
\sigma^{(cl)}_{q}  = \frac{{\langle q^2 \rangle}_{(cl)}}{\hbar/(2m\omega_0)}
= \frac{2}{\beta\omega_0} +
\frac{2\omega_0}{\pi\omega_c} \, \Psi \left(1 \right)
\sum_{i=1}^{4}  A_i   = \frac{2}{\beta\omega_0} 
\, 
\longrightarrow
\,  {\langle q^2 \rangle}_{(cl)} = \frac{k_BT}{m\omega_0^2} \,\,  ,
\end{equation}
in which we used $\sum_{i=1}^{4} A_i=0$. 
Going further in the high temperature expansion, 
we can obtain quantum corrections to the classical result which are proportional to the thermal de Broglie wavelength
%
%
%
%
%
%
\begin{equation}
\label{eq:hi_T}
{\langle q^2 \rangle}_{high-T} = {\langle q^2 \rangle}_{(cl)} + \frac{q_0^2}{6} \beta\omega_0 \left( 1 + \frac{\gamma_p\omega_c}{\omega_0^2}\right)
\end{equation}
with $q_0^2  \beta\omega_0=\hbar^2/(2mk_BT)$.  
Here we used $\sum_{i=1}^{4} A_i \Omega_i =  \omega_c(1+\gamma_p\omega_c/\omega_0^2)$. 
%
%
Notice that, even if the temperature is relatively high $\beta\omega_0  \lesssim  1$,  quantum corrections to the fluctuations can become relevant 
in presence of the interaction with a second bath via the momentum operator for $\gamma_p \omega_c/\omega_0^2 \gg 1$.
Although the result depends on the choice of the spectrum for the response function 
(in this case of a Drude form with a high-frequency cutoff), 
one can expect that the bath coupled through the operator $\hat{p}$ of the oscillator acts as additional source of quantum noise for 
the operator $\hat{q}$. 
The result (\ref{eq:hi_T}) represents the dual expression of the standard, damped harmonic oscillator with Ohmic-Drude dissipation 
for which we have ${\langle p^2 \rangle}_{high-T} \simeq {\langle p^2 \rangle}_{(cl)} +  m \gamma_q \omega_c \hbar^2/(12k_BT)$ \cite{Weiss:2012} 
in the limit $\omega_c \gg (\omega_0,\gamma_q)$.

In the opposite, low temperature regime, we consider the expansion for the digamma function $\Psi(1+x)$ for $x \gg 1$, 
which implies $\beta \Omega_i /(2\pi) \gg 1$ for $i=1,\dots,4$.
We then obtain quadratic corrections in  $T$ 
%
%
%
%
%
%
\begin{eqnarray}
\sigma^{(low-T)}_{q}  &=& 
\frac{2}{\beta\omega_0} +
\frac{2\omega_0}{\pi\omega_c} \, 
\sum_{i=1}^{4}  A_i  
\left[  \log\left(  \frac{\beta \Omega_i}{2\pi}\right) + \frac{1}{2}\left(\frac{2\pi}{\beta \Omega_i}\right) - \frac{1}{12} 
{\left(\frac{2\pi}{\beta \Omega_i}\right)}^2
\right]  \nonumber \\
&=& \sigma^{0}_{q}  \,\, +  \frac{2\pi}{3} \left(  \frac{\gamma_q}{\omega_0} \right) {\left( \frac{k_BT}{\hbar\omega_0} \right)}^2  \, .
\end{eqnarray}
The linear term in $T$ cancels with the first term due to $\sum_{i=1}^{4}  A_i/\Omega_i = -\omega_c/\omega_0^2$.
We have also used  $\sum_{i=1}^{4}  A_i/\Omega_i^2=-\gamma_q\omega_c/\omega_0^4$. 
Thus, for sufficiently low temperature $T \ll T^*_q$, defined by
%
%
%
%
%
%
\begin{equation}
\label{eq:T_c} 
T^*_q = \mbox{min }\left[ \{ \hbar |\Omega_i|  \} ,  \sqrt{ \frac{3\omega_0}{2\pi\gamma_q}  }  \hbar\omega_0 \right] \quad (i=1,\dots,4) \, ,
\end{equation}
we can neglect the finite temperature effects for the fluctuations of the position operator $\hat{q}$.
The zero temperature limit of the $\hat{q}$ fluctuations reads 
%
%
%
%
%
%
\begin{equation}
\label{eq:q2_0} 
\sigma^{0}_{q}  =
\frac{2\omega_0}{\pi\omega_c} \left[
\log\left(  \frac{\beta \omega_c}{2\pi}\right) \sum_{i=1}^{4}  A_i  
+ \sum_{i=1}^{4}  A_i  \log\left(  \frac{\Omega_i}{\omega_c}\right) 
\right] =  \frac{2\omega_0}{\pi\omega_c} \sum_{i=1}^{4}  A_i  \log\left(  \frac{\Omega_i}{\omega_c}\right) 
\end{equation}
where we used again $\sum_{i=1}^{4}  A_i=0$. 
By interchanging the damping coefficients $\gamma_q \leftrightarrow \gamma_p$, a 
similar expression to (\ref{eq:T_c})  and (\ref{eq:q2_0}) hold for the temperature threshold  $T^*_p$  for  the quantum regime 
and for the quantum fluctuations $\sigma^{0}_{p} $ of the momentum operator $\hat{p}$.
In the following we concentrate on the behaviour of the quantum fluctuations.

\section{Zero temperature fluctuations}
\label{sec:T=0}

In this section, assuming the limit of low temperature, we will use the 
result determined in Eq.~(\ref{eq:q2_0})   to discuss the ground state fluctuations in the different regimes.
For the sake of completeness, we recall the regime of squeezing of the oscillator 
in which we have the case $\langle \hat{Q}^2 \rangle < 1$ {\sl or}  $\langle \hat{P}^2 \rangle < 1$, 
and we discuss in detail the enhancement of the quantum fluctuations, 
for example $\langle \hat{Q}^2 \rangle  \gg 1$ {\sl and} $\langle \hat{P}^2 \rangle > 1$. 
The cross-over between these two regimes is also analyzed.

We consider the low-frequency expansion $\omega_0,\gamma_{q} ,\gamma_{p} \ll \omega_c$ for which we can find a simple analytic expression for the
frequencies $\Omega_i$. 
In this way, one obtains an analytic expansion for the roots of the quartic polynomial (\ref{eq:P4}) and hence for the frequencies $\Omega_i$ 
related to the quantum fluctuations. 
Note that the frequencies $\Omega_i$ are related to the poles $z_i$ of the denominator 
(\ref{eq:Den}) as $z_i = - i \Omega_i$.
As the real parts of the frequencies  $\Omega_i$  have the same sign, this implies that 
the imaginary parts of $z_i$ have also the same sign, as assumed in the previous section.
 
First of all, we discuss the limit in which the results for a single bath are recovered \cite{Weiss:2012}. 
This limit is defined by $\gamma_p \ll {(\omega_0/\omega_c)}^2  \gamma_q/4$ (viz. the bath coupled to the oscillator 
via the position $\hat{q}$ dominates) or equivalently by  $\gamma_q \ll {(\omega_0/\omega_c)}^2  \gamma_p/4$ 
(viz. the bath coupled to the oscillator via the position $\hat{p}$ dominates). 
In this case we obtain as a solution for the expansion
%
%
%
%
%
%
 \begin{equation}
\label{eq:Om_single_bath} 
\Omega_{1,2} = \frac{\mbox{max}\{\gamma_p,\gamma_q\}}{2}  \pm  i \sqrt{ \omega_0^2- {\left(\frac{\mbox{max}\{\gamma_p,\gamma_q\}}{2}\right)}^2  } \, , \,\,\,\,
\Omega_{3}  =  \omega_c - \mbox{max}\{\gamma_p,\gamma_q\}   \, , \,\,\,\,
\Omega_{4}  =  \omega_c  \,.
\end{equation}
Since one frequency equals the cutoff $\Omega_{4}  =  \omega_c$, the coefficient $A_4=0$ and the sum Eq.~(\ref{eq:q2_0})  reduces only to 
three terms as in the case of the damped harmonic oscillator with the Drude regularization  \cite{Weiss:2012}. 
Along this line, it is possible to show that, from Eq.~(\ref{eq:q2_0}), one recovers the known results for the
fluctuations of the damped harmonic oscillator in the cases $\gamma_p=0$ \cite{Weiss:2012} 
or  $\gamma_q=0$  \cite{Cuccoli:2001,Ankerhold:2007}, 
taking into account the symmetry $\left< Q^2 \right> = \sigma(\gamma_q,\gamma_p)$ and $\left< P^2 \right> = \sigma(\gamma_p,\gamma_q)$.
In this regime,  the quantum fluctuations of the quadrature coupled to the bath are squeezed and the ones of the conjugate variable are enhanced.

Far away from the single bath regime, one obtains the following results for the low-frequency expansion
%
%
%
%
%
%
\begin{equation}
\Omega_{1,2} =
\frac{\omega_0}{1+\rho^2} \left[  \Gamma  \pm i \sqrt{ 1 - \Delta\Gamma_{q,p}^2 }  \right]   
\, , \quad 
\Omega_{3,4} = \omega_c \left(1 \pm i \rho \right) -   \frac{\omega_0 \Gamma}{1 \pm i \rho}  \, ,
\label{eq:Om_general}
\end{equation}
in which we set  $\rho=\sqrt{\gamma_q\gamma_p}/\omega_0$, $\Gamma = (\gamma_q+\gamma_p)/(2\omega_0)$ and 
$\Delta\Gamma_{q,p} = (\gamma_q-\gamma_p)/(2\omega_0)$.
From this result, it is clear that in the regime $|\gamma_q-\gamma_q| < 2\omega_0$, viz. $|\Delta\Gamma_{q,p}|<1$, 
all frequencies are complex and it is possible to show that 
the relaxation dynamics of the harmonic oscillator is always underdamped, i.e. 
the dynamical correlation functions exhibit always  an oscillating decay 
even for large damping $(\gamma_q,\gamma_p)\gg\omega_0$  \cite{Kohler:2006}.

\begin{flushleft}
\begin{figure}[tb]
\begin{center}
\includegraphics[scale=0.35,angle=-90.]{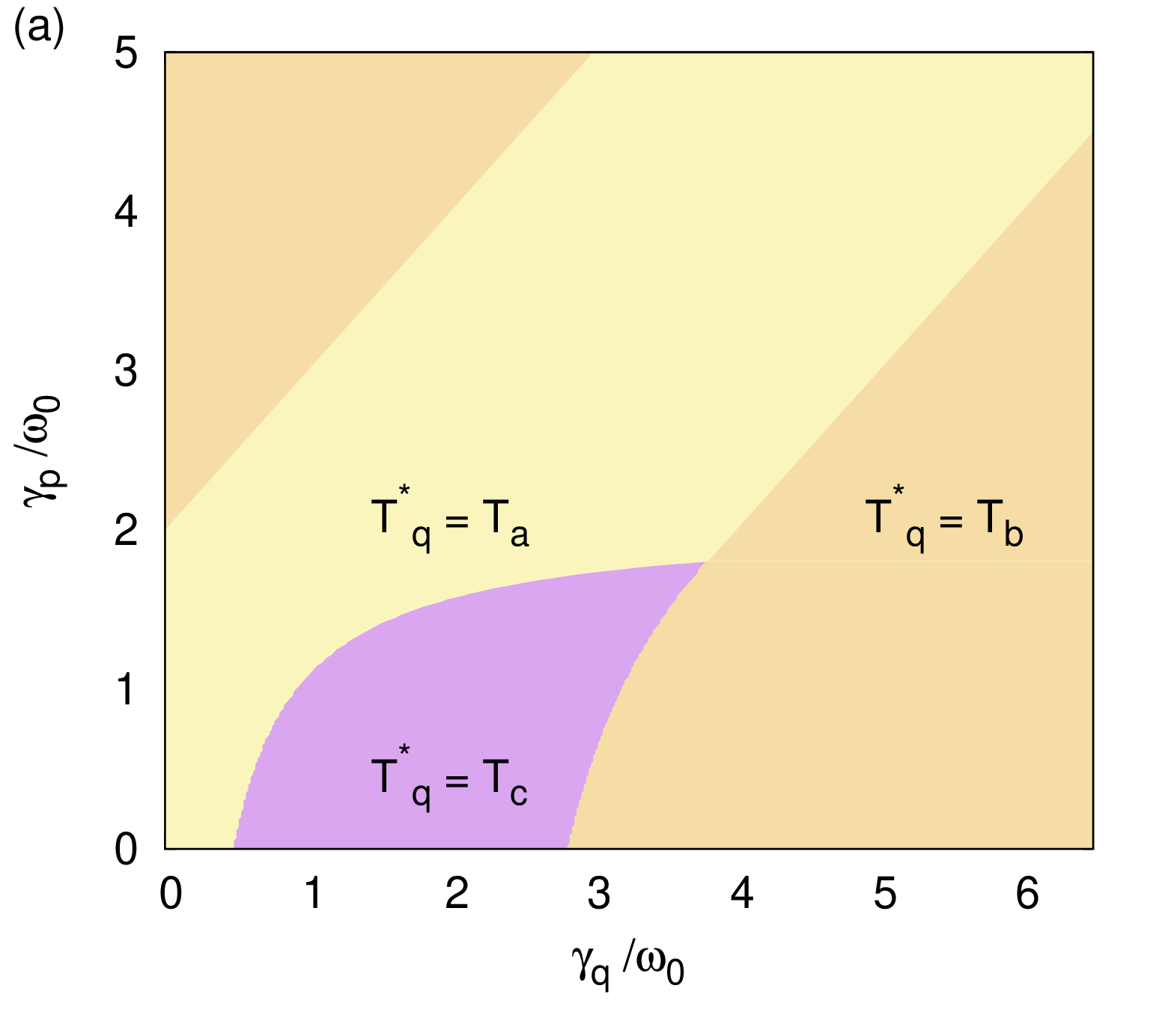}\includegraphics[scale=0.35,angle=-90.]{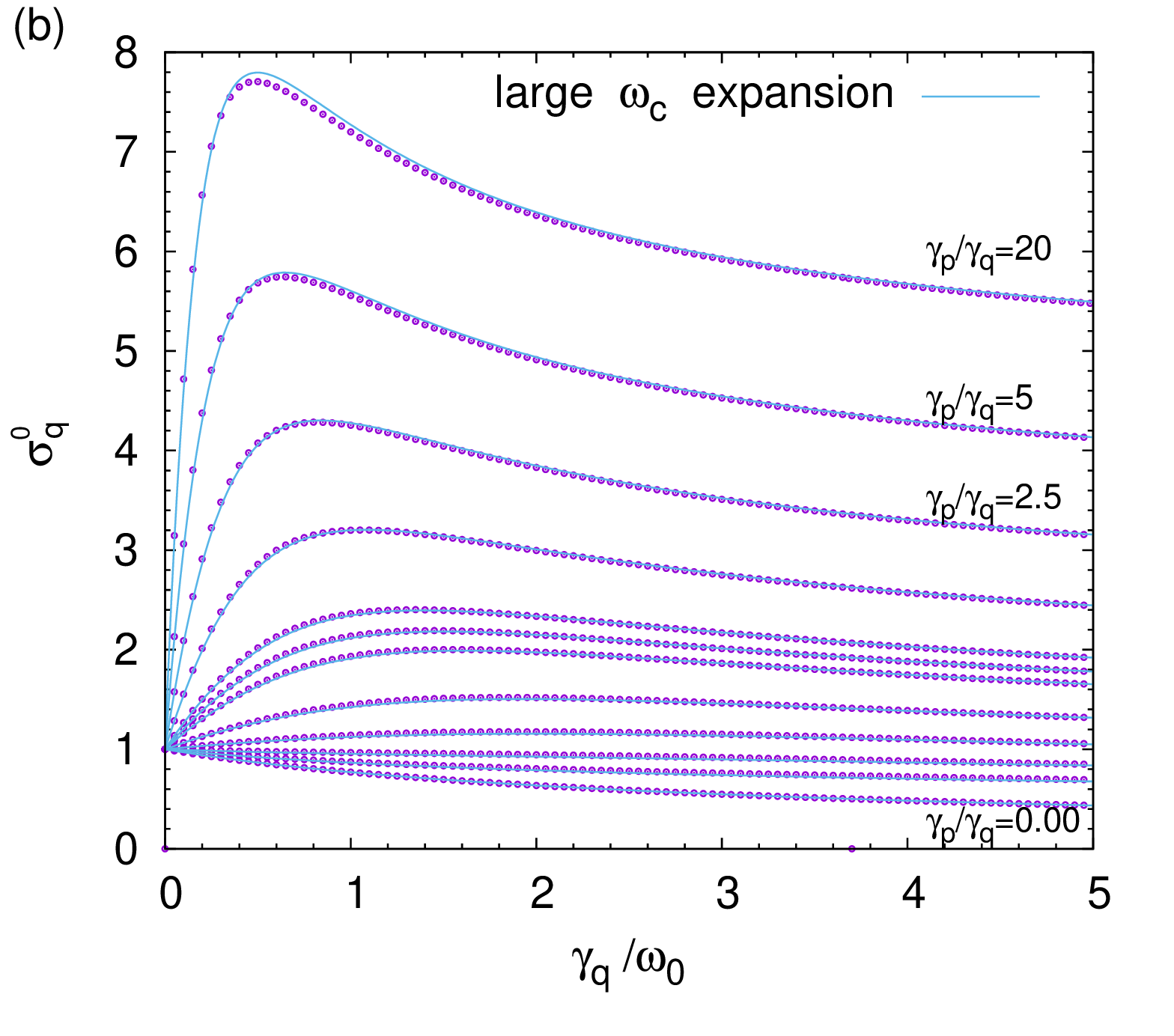}
\end{center}
\caption{
(a) Temperature threshold $T^*_q$ for the quantum regime of the fluctuations of $\hat{q}$ defined for the large cutoff expansion,  
see Eq.~(\ref{eq:T_c_wc}) for the definition of $T_a,T_b,T_c$.
(b) Comparison between Eq. (\ref{eq:q2_0}) for the quantum fluctuations (dots) and the large cutoff $\omega_c$ expansion 
Eq.~(\ref{eq:u2_general_Hif})  (solid line)  for $\omega_c/\omega_0=80$. 
The different curves correspond to different ratios $\gamma_p/\gamma_q=20, 10, 5, 2.5, 1.25, 1, 0.8, 0.4, 0.2, 0.1, 0$ (from top to bottom).
}
\label{fig:1}
\end{figure}
\end{flushleft}

From the analytic expression of $\Omega_i$ we infer the temperature threshold $T^*_q$ for the quantum regime defined in Eq.~(\ref{eq:T_c}) 
for the fluctuations of $\hat{q}$.
The result is shown in Fig.\ref{fig:1}(a).
Using the expansion  Eqs.~(\ref{eq:Om_general}) 
for the frequencies $\{ \Omega_i \}$ (i=1,\dots,4),  the three temperatures shown in Fig.\ref{fig:1}(a)
correspond  to
%
%
%
%
%
%
\begin{equation}
k_BT_a= \frac{\hbar\omega_0}{\sqrt{1+\rho^2} } \, , \quad
k_BT_b=\hbar\omega_0 \frac{\Gamma - \sqrt{ \Delta\Gamma^2_{q,p}-1}}{1+\rho^2} \, , \quad
k_BT_c=\hbar\omega_0 \sqrt{\frac{3\omega_0}{2\pi\gamma_q}} \,\, .
\label{eq:T_c_wc}
\end{equation}
%
%
%
%
%
%
%
%
%
%
%
%
%
%
%
Using the same expansion in $\omega_0,\gamma_{q} ,\gamma_{p} \ll \omega_c$ for the expression of the quantum fluctuations (\ref{eq:q2_0}) 
and of the coefficients (\ref{eq:A_coeff}), 
we finally  obtain the following analytic expression for the zero temperature fluctuations
%
%
%
%
%
%
\begin{equation}
\label{eq:u2_general_Hif} 
\tilde{\sigma}^{0}_{q}  
=
\frac{2}{\pi(1+\rho^2)}
\left[
\frac{\gamma_p}{\omega_0} \left( 
\ln\left( \frac{\omega_c}{\omega_0}\right)  
+
\rho \arctan(\rho)+ \ln(1+\rho^2)   
\right)
+
\frac{
\left(
1 + \frac{\gamma_p}{\omega_0} \Delta\Gamma_{q,p}
\right)
}{\sqrt{\left|1 -  \Delta\Gamma_{q,p}^2\right|}  } \Theta_{q,p} 
\right]
\end{equation}
with
%
%
%
%
%
%
%
\begin{equation}
\label{eq:F_chi} 
 \Theta_{q,p} 
=
\left\{
\begin{array}{cc}
\arctan   \left(  \sqrt{1-\Delta\Gamma_{q,p}^2}  \, /  \, \Gamma \right)  & \mbox{for} \qquad \left| \Delta\Gamma_{q,p} \right| < 1 \\
\arctanh \left( \sqrt{\Delta\Gamma_{q,p}^2-1} \, /  \, \Gamma  \right) & \mbox{for} \qquad \left| \Delta\Gamma_{q,p} \right| > 1
\end{array}
\right. \, .
\end{equation}
In Fig.\ref{fig:1}(b) we show the results of the comparison between the 
exact formula $\sigma_q^0$ for the quantum fluctuations (\ref{eq:q2_0})  and the large $\omega_c$  
expansion $\tilde{\sigma}^{0}_{q}$ Eq.~(\ref{eq:u2_general_Hif}). 
We observe that the large $\omega_c$ expression is in excellent agreement with the 
exact formula almost all values of the ratio between the coupling 
strenghts of the two baths $\gamma_p/\gamma_q$, both in the underdamped $\gamma_{q,p} < \omega_0$ 
and in the overdamped regime $\gamma_{q,p} > \omega_0$.   
When the bath coupled to the position dominates $\gamma_q \gg \gamma_p$,
we are in the limit of a single bath and the flucuations are squeezed  
with increasing dissipative coupling $\gamma_q$. 
Moreover, fixing $\gamma_q$, the fluctuations of $\hat{q}$ increase with larger $\gamma_p$, viz. the coupling strength 
of the conjugate variable $\hat{p}$, as follows from Eq.~(\ref{eq:u2_general_Hif}). 
Nevertheless the surface $\sigma_{q}$ as a function of $(\gamma_q,\gamma_p)$ displays a non-trivial behavior 
which can be seen by considering this function along lines of constant  ratio $\gamma_p/\gamma_q$, 
as shown in Fig.\ref{fig:1}(b).
In this case, the fluctuations can show a {\it non-monotic} behavior at large ratios $\gamma_p/\gamma_q$. 
This result was obtained numerically in Ref.\cite{Cuccoli:2010} , while here we provide an analytic derivation.
By inspection of the analytic expression, the initially increasing slope of the fluctuations 
is strongly determined by the first linear term of Eq.~(\ref{eq:u2_general_Hif})
which is proportional to the logarithm of the large frequency cutoff. 
Therefore we conclude that this behavior is sensible to the high-frequency part of the bath's spectrum.
%
%
%
%
%
%
%
\begin{flushleft}
\begin{figure}[tbhp]
\begin{center}
\includegraphics[scale=0.35,angle=-90.]{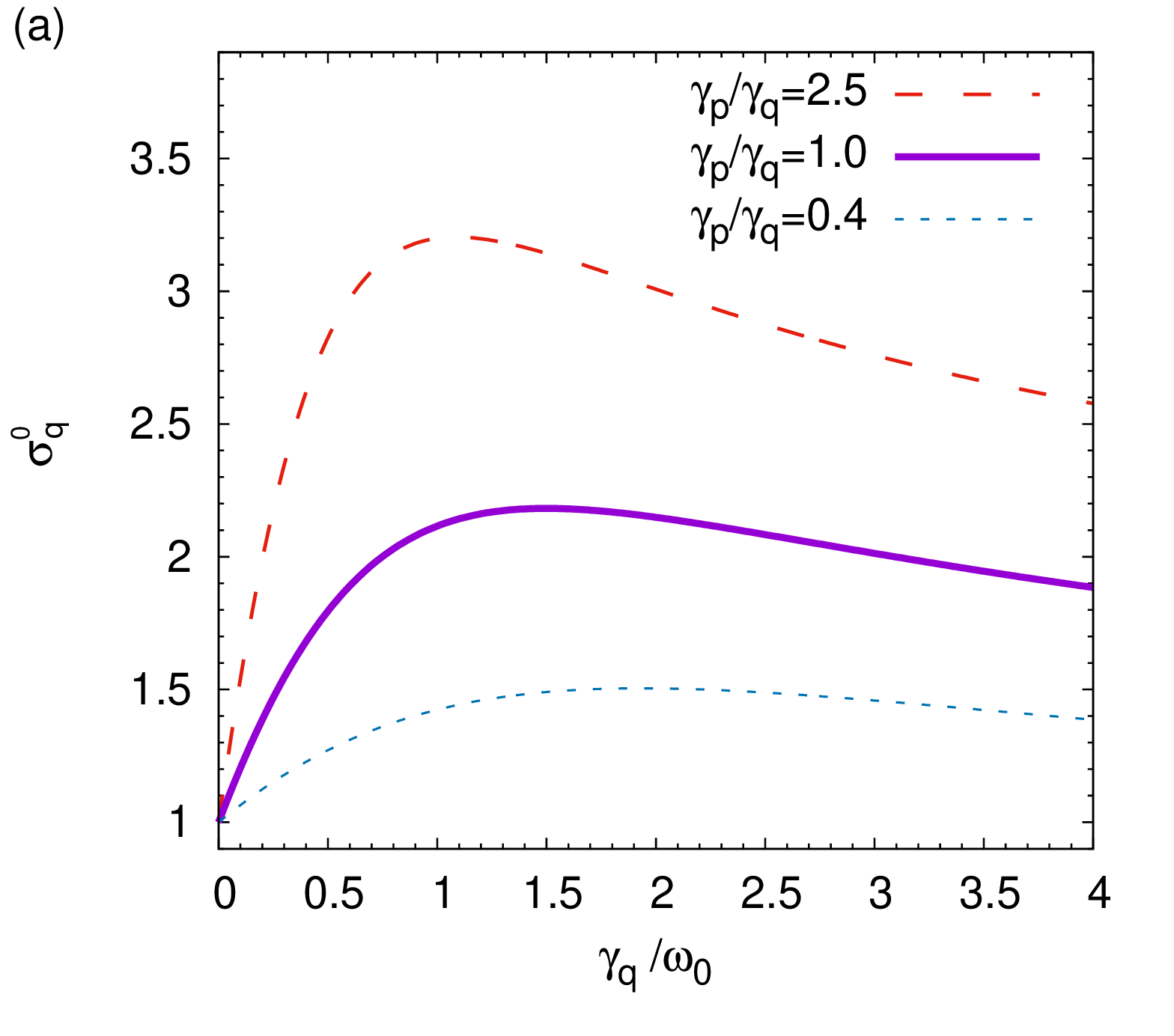}\includegraphics[scale=0.35,angle=-90.]{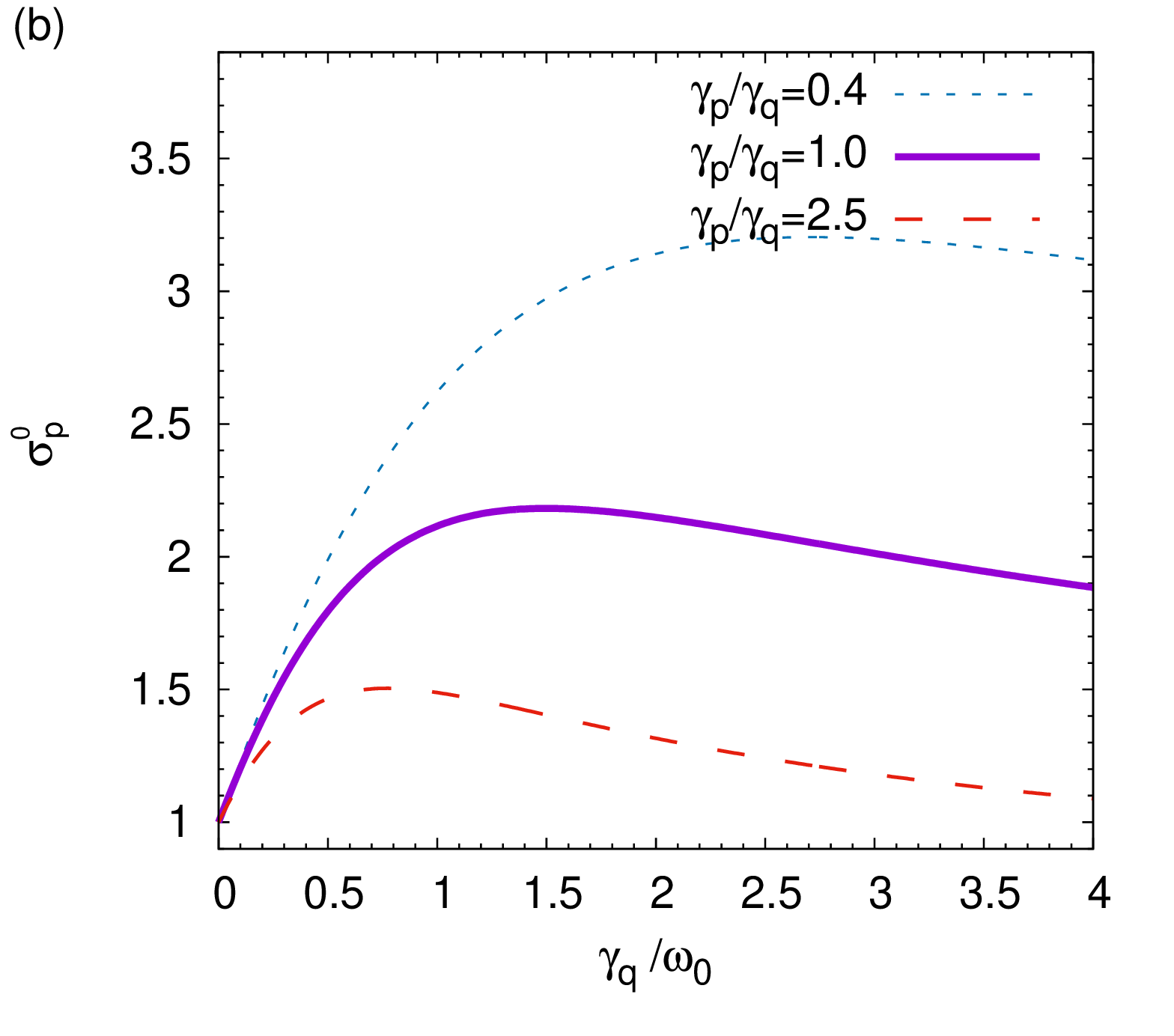}
\end{center}
\caption{
Scaled zero-temperature quantum fluctuations of (a) $\hat{q} $  and (b)  $\hat{p} $ for the large cutoff expansion and for 
different curves corresponding to three different ratios $\gamma_p/\gamma_q=2.5, 1, 0.4$.
The cutoff is $\omega_c/\omega_0=80$.
}
\label{fig:2}
\end{figure}
\end{flushleft}
%
%
%
We underline  that the regime of {\sl enhancement } of quantum fluctuations corresponds to the case when {\sl both} fluctuations of 
$\hat{q}$ and $\hat{p}$ grow with increasing the dissipative coupling constant $\gamma_{q}$ and $\gamma_{p}$.
An example is shown in Fig.~\ref{fig:2}.
We note that the curve $\gamma_q=\gamma_p$ is identical for the position and momentum fluctuations whereas 
the other curves appear different as we plot the fluctuations as a function of the parameter $\gamma_q$ only.

Interestingly, in the intermediate range of damping defined by $(\gamma_{q},\gamma_{p}) \sim \omega$,  
it is possible to reach a strong enhancement of the quantum fluctuations.
For example, as shown in Fig.~\ref{fig:2}(a), the quantum fluctuations of the position $\sigma_q^0$ 
are larger than twice the bare quantum fluctuations  at  $\gamma_q = 0.5 \omega_0$ and $\gamma_p =1.25 \omega_0$.
In the same range, we also observe substantial squeezing of the fluctuations.
For example, in  Fig.~\ref{fig:1}(b), at $\gamma_p \ll (\gamma_q,\omega_0)$, we are in the regime of a single bath and 
the fluctuations $\sigma_q^0$ are squeezed by a factor $\sim 0.6$ at $\gamma_q=\omega_0$.
We observe that this intermediate range of damping $(\gamma_{q},\gamma_p) \sim \omega$ 
corresponds to a temperature threshold which is of order of $\omega_0$.
In other words, the condition for low temperature can be simplified, roughly speaking, as $k_B T \ll \hbar \omega_o$ .

\section{Summary and perspectives}
\label{sec:conclusions}

We studied the fluctuations of the harmonic oscillator coupled to two independent baths
via the two conjugate variables, viz. the position $\hat{q}$ and the momentum $\hat{p}$.
%
%
For the Ohmic-Drude dissipation, we derived analytic formulas for the fluctuations in the high and low temperature limit.
Importantly, we calculated the temperature threshold $T^*$ below which 
quantum fluctuations represent the dominant contribution and finite temperature corrections are negligible.
We analyzed the enhancement and the squeezing of the quantum fluctuations as varying the damping coefficients 
$\gamma_q$ and $\gamma_p$ respect to the oscillator's frequency $\omega_0$.
In the intermediate damping regime $\gamma_{\nu} \sim \omega_0$, such effects are significant and detectable, 
provided that the oscillator can be cooled to low temperature $T \ll T^*$  with  $k_B T^* \sim \hbar \omega_0$.
%
%
%
%
%
%
%
%
%
%
%
%
\begin{flushleft}
\begin{figure}[tb]
\begin{center}
\includegraphics[scale=0.7,angle=0.]{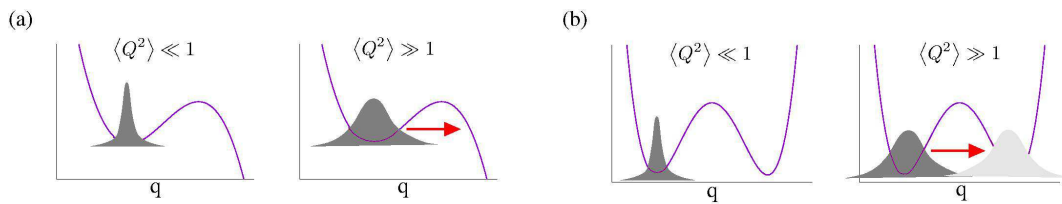}
\end{center}
\caption{Schematic picture of (a) metastable state and (b) double well potential $V(q)$.
Assuming a strong conventional dissipative coupling with a first bath via the position $q$, 
the scaled quantum fluctuations are squeezed $\left< Q^2 \right> \ll 1$ and quantum tunnelling is quenched.
Coupling the system to a second bath via the momentum operator $p$ can yield enhancement of  the quantum fluctuations 
$\left< Q^2 \right> \gg 1$, eventually restoring tunnelling.}
\label{fig:3}
\end{figure}
\end{flushleft}

The enhancement of quantum fluctuations can be useful to achieve quantum effects in systems for which this issue is still an 
open challenge. \\
For instance, one can consider a potential $V(q)$ with a local minimum (metastable state) or a double-well potential, see Fig.~\ref{fig:3}, in which  
the harmonic oscillator states are localized around the minima. 
One can assume the case in which such states are even strongly localized due to the standard dissipative coupling - via the position $q$ -  
such that quantum tunnelling is quenched, e.g. the so-called ``localized phase" \cite{Weiss:2012}.
For these systems, one can engineer a coupling with a second bath via the momentum $p$ such that quantum fluctuations are enhanced,  
eventually restoring quantum tunnelling and hence the quantum delocalized phase in the system.\\
Indeed, the observation of quantum macroscopic tunnelling in opto-mechanical and electro-mechanical systems can be a difficult task 
as these systems are particularly massive and, hence, they are generally in a regime in which quantum tunnelling is undetectable.
Coupling the (non-linear) mechanical oscillator to a second bath via its momentum can lead to an enhancement 
of its quantum fluctuations opening the possibility of observing macroscopic quantum tunneling effects even in such systems.  
Future works will explore this perspective in real nanomechanical devices.


\ack
 
G.R. acknowledges S. Andergassen and W. Belzig  for a careful reading of the manuscript
and E. Buks for an interesting discussion.
This research was supported by the Zukunftskolleg of the University of Konstanz, by the DFG through SFB 767 and 
by the MWK-RiSC program, project N.13551400.


%

\section*{References}

\bibliographystyle{iopart-num}
\bibliography{references}

\providecommand{\newblock}{}
\begin{thebibliography}{10}
\expandafter\ifx\csname url\endcsname\relax
  \def\url#1{{\tt #1}}\fi
\expandafter\ifx\csname urlprefix\endcsname\relax\def\urlprefix{URL }\fi
\providecommand{\eprint}[2][]{\url{#2}}

\bibitem{Weiss:2012}
Weiss U 2012 {\em Quantum Dissipative Systems\/} 4th ed (Singapore: World
  Scientific Publishing)

\bibitem{Breuer:2012}
Breuer H~P and Petruccione F 2007 {\em The Theory of Open Quantum Systems\/}
  2nd ed (Oxford University Press)

\bibitem{Wiseman:2014}
Wiseman H~M and Milburn G~J 2014 {\em Quantum Measurement and Control\/} 2nd ed
  (Cambridge University Press)

\bibitem{Zagoskin:2011}
Zagoskin A~M 2011 {\em Quantum Engineering: theory and design of quantum
  coherent structures\/} (Cambridge University Press)

\bibitem{Schlosshauer:2010}
Schlosshauer M~A 2010 {\em Decoherence and the Quantum-To-Classical
  Transition\/} (Springer)

\bibitem{Leggett:2002}
Leggett A~J 2002 {\em Journal of Physics: Condensed Matter\/} {\bf 14}
  R415--R451

\bibitem{Raimond:2001}
Raimond J~M, Brune M and Haroche S 2001 {\em Rev. Mod. Phys.\/} {\bf 73}(3)
  565--582

\bibitem{Haroche:2013}
Haroche S 2013 {\em Rev. Mod. Phys.\/} {\bf 85} 1083--1102

\bibitem{Wineland:2013}
Wineland D~J 2013 {\em Rev. Mod. Phys.\/} {\bf 85} 1103--1114

\bibitem{Bassi:2013}
Bassi A, Lochan K, Satin S, Singh T~P and Ulbricht H 2013 {\em Rev. Mod.
  Phys.\/} {\bf 85} 471--527

\bibitem{DeMartini:2012}
De~Martini F and Sciarrino F 2012 {\em Rev. Mod. Phys.\/} {\bf 84} 1765--1789

\bibitem{Nielsen:2012}
Nielsen M~A and Chuang I~L 2011 {\em Quantum Computation and Quantum
  Information\/} 10th ed (Cambridge University Press)

\bibitem{Weiss:1985-2}
Riseborough P~S, H{\"a}nggi P and Weiss U 1985 {\em Phys. Rev. A\/} {\bf 31}
  471--478

\bibitem{Grabert:1988}
Grabert H, Schramm P and G-L I 1988 {\em Phys. Rep\/} {\bf 168} 115--207

\bibitem{Unruh:1989}
Unruh W~G and Zurek W~H 1989 {\em Phys. Rev. D\/} {\bf 40} 1071--1094

\bibitem{Girvin:2009}
Girvin S~M, Devoret M~H and Schoelkopf R~J 2009 {\em Physica Scripta\/} {\bf
  2009} 014012

\bibitem{Poot:2012}
Poot M and van~der Zant H~S~J 2012 {\em Physics Reports\/} {\bf 511} 273--335

\bibitem{Aspelmeyer:2014}
Aspelmeyer M, Kippenberg T~J and Marquardt F 2014 {\em Rev. Mod. Phys.\/} {\bf
  86} 1391--1452

\bibitem{Xiang:2013}
Xiang Z~L, Ashhab S, You J~Q and Nori F 2013 {\em Rev. Mod. Phys.\/} {\bf 85}
  623--653

\bibitem{Dykman:2012}
Dykman M 2012 {\em Fluctuating Nonlinear Oscillators: From Nanomechanics to
  Quantum Superconducting Circuits\/} 1st ed (Oxford University Press)

\bibitem{Cuccoli:2001}
Cuccoli A, Fubini A, Tognetti V and Vaia R 2001 {\em Physical Review E\/} {\bf
  64} 066124

\bibitem{Ankerhold:2007}
Ankerhold J and Pollak E 2007 {\em Physical Review E\/} {\bf 75} 041103

\bibitem{Kohler:2005}
Kohler H and Sols F 2005 {\em Physical Review B\/} {\bf 72} 180404

\bibitem{Kohler:2006}
Kohler H and Sols F 2006 {\em New Journal of Physics\/} {\bf 8} 149--149

\bibitem{Cuccoli:2010}
Cuccoli A, Del~Sette N and Vaia R 2010 {\em Physical Review E\/} {\bf 81}
  041110

\bibitem{Kohler:2013-2}
Kohler H and Sols F 2013 {\em Physica A: Statistical Mechanics and its
  Applications\/} {\bf 392} 1989--1993

\bibitem{Neto:2003}
Neto A~H~C, Novais E, Borda L, Zarand G and Affleck I 2003 {\em Physical review
  letters\/} {\bf 91} 096401

\bibitem{Novais:2005}
Novais E, Castro~Neto A~H, Borda L, Affleck I and Zarand G 2005 {\em Physical
  Review B\/} {\bf 72} 014417

\bibitem{Kohler:2013}
Kohler H, Hackl A and Kehrein S 2013 {\em Physical Review B\/} {\bf 88} 205122

\bibitem{Bruognolo:2014}
Bruognolo B, Weichselbaum A, Guo C, von Delft J, Schneider I and Vojta M 2014
  {\em Physical Review B\/} {\bf 90} 245130

\bibitem{Zhou:2015}
Zhou N, Chen L, Xu D, Chernyak V and Zhao Y 2015 {\em Physical Review B\/} {\bf
  91} 195129

\bibitem{Lang:2015}
Lang N and B{\"u}chler H~P 2015 {\em Physical Review A\/} {\bf 92} 012128

\end{thebibliography}

\end{document}